\pdfoutput=1
\documentclass[letterpaper,twocolumn,10pt]{article}
\usepackage{amsmath}%
\usepackage{grffile}
\usepackage{graphicx}
\usepackage{url}
\usepackage{subfig}
\usepackage{epstopdf}
\usepackage{amsfonts}%
\usepackage{amssymb}%
\usepackage{array,multirow}
\usepackage{fancyvrb}
\usepackage{fancybox}
\usepackage[utf8]{inputenc}
\usepackage{xcolor}
\usepackage[linesnumbered,ruled]{algorithm2e}
\usepackage[margin=1in,footskip=0.25in]{geometry}
\usepackage{listings}
\usepackage{array}
\usepackage{usenix}

\SetAlFnt{\small}
\SetAlCapFnt{\normalsize}
\SetAlCapNameFnt{\normalsize}

\newenvironment{packed_itemize}{
\begin{itemize}
  \setlength{\itemsep}{2pt}
  \setlength{\parskip}{1pt}
  \setlength{\parsep}{0pt}
}{\end{itemize}}

\makeatletter
\newenvironment{CenteredBox}{%
\begin{Sbox}}{
\end{Sbox}\centerline{\parbox{\wd\@Sbox}{\TheSbox}}}
\makeatother
\begin{document}

\title{Did we learn from LLC Side Channel Attacks? 
A Cache Leakage Detection Tool for Crypto Libraries}

\author{
	{\rm Gorka Irazoqui\textsuperscript{$*$}, Kai Cong\textsuperscript{$\dagger$}, Xiaofei Guo\textsuperscript{$\zeta$}, Hareesh Khattri\textsuperscript{$\dagger$},} \\
	{\rm Arun Kanuparthi\textsuperscript{$\dagger$}, Thomas Eisenbarth\textsuperscript{$*$}, and Berk Sunar\textsuperscript{$*$}}\\
	\textsuperscript{$*$}Worcester Polytechnic Institute, \textsuperscript{$\zeta$}Tetration Analytics, \textsuperscript{$\dagger$}Intel
	}

\maketitle

\begin{abstract}

This work presents a new tool to verify the correctness of cryptographic implementations with respect to cache attacks. Our methodology 
 discovers vulnerabilities that are hard to find with other techniques, observed as exploitable leakage. The methodology works by identifying secret dependent memory and introducing forced evictions inside potentially vulnerable code to obtain cache traces that are analyzed using Mutual Information. If dependence is observed, the cryptographic implementation is classified as to leak information.

We demonstrate the viability of our technique in the design of the three main cryptographic primitives, i.e., AES, RSA and ECC, in eight popular up to date cryptographic libraries, including OpenSSL, Libgcrypt, Intel IPP and NSS. Our results show that cryptographic code designers are far away from incorporating the appropriate countermeasures to avoid cache leakages, as we found that 50\% of the default implementations analyzed leaked information that lead to key extraction. We responsibly notified the designers of all the leakages found and suggested patches to solve these vulnerabilities.
\end{abstract}

\section{Introduction}

In the last decade we have witnessed the cloud revolution enabled by sandboxing mechanisms such as virtualization.
These technologies allow cloud service providers to place data and processes of various customers on the same hardware without jeopardizing their security.
IaaS clouds for instance rent expensive hardware resource to multiple customers by offering guest OS instances sandboxed inside virtualized machines (VMs). PaaS cloud services go one step further and allow users to share the application space while sandboxed, e.g. using containers, at the OS level. Similar sandboxing techniques are used to isolate semi-trusted apps running on mobile devices. Even browsers use sandboxing to execute untrusted code without the risk of harming the local host.

Resource sharing improves utilization and thereby helps reducing IT costs and saves power. However, resource sharing also enables information leakages at the hardware level which sandboxing techniques cannot prevent and that can be exploited by malicious code. One of the most prominent leakage sources, i.e. the Last Level Cache (LLC), has been heavily targeted. From the adversaries' point of view, the LLC leakage channel has a number of advantages, including high resolution information and cross core information leakage.
 LLC attacks have demonstrated to recover a wide range of information, ranging from private cryptographic keys~\cite{184415} to user's private shopping behavior~\cite{Zhang:2014:CSA:2660267.2660356}. Furthermore, LLC attacks have been successfully applied in a wide range of scenarios, e.g., malicious VMs in IaaS or PaaS public clouds, malicious Javascript execution in web browsers~\cite{DBLP:journals/corr/OrenKSK15} or even as malicious smartphone apps~\cite{DBLP:conf/uss/LippGSMM16}. In short, cache attacks have demonstrated to pose a severe threat applicable in a wide range of use scenarios. 

These attacks can be stopped at three different layers; at the hardware~\cite{Wang:2007:NCD:1250662.1250723}, the OS/hypervisor ~\cite{Liu_GYMRHL_16} or at the application layer~\cite{andi1}. While the first two involve overheads that hardware/OS designers might not be willing to pay, preventing leakage in security critical applications can be achieved by aware code designers through proper implementation techniques.
Indeed, cache attacks usually exploit human coding mistakes that lead to private information leakage. Although these leakages might sometimes be difficult to detect, they need to be carefully verified, especially when designing cryptographic code. 
Cryptographic primitives make the core of the security infrastructure used to protect our digital communications. If the used cryptographic code is poorly designed, any solution built on top is also susceptible to fail.

In summary, cache attacks diverge from traditional attacks in the following ways:
\begin{packed_itemize}
\item {\bf Stealthy:} Leakage attacks are extremely hard to detect. The effects of an attack may only be felt through performance degradation for a short duration while leakage traces are being collected. After the attack is performed no digital footprint is left. Hence standard detection techniques such as traffic, access and privilege monitoring are completely blind to leakage attacks. For instance, any application submitted to the app store is checked for access violations to devices, however, any attack code exploiting hardware leakage would only monitor legitimate memory access time variations.
 
\item {\bf Hard to Prevent:} It is rather difficult to design leakage proof code, especially if good performance is also an objective. Leakages are quite subtle and may not be detected after deployment for a long time. Even if the code is not leaky on one platform, with gradual optimizations implemented at the microarchitectural level, new leakage channels may emerge with newer releases of a platform.

\item {\bf Difficult to Test:} Verification of leakage resistant code is painfully slow and typically requires inspection by experts well versed in cache attacks. 
\end{packed_itemize}

In this paper, we introduce a tool that helps cryptographic code designers to analyze and expose any leakages in cryptographic implementations. We demonstrate the viability of the technique by analyzing cache leakages in a number of popular cryptographic libraries with respect to three main cryptographic primitives we rely on every day to securely communicate over the internet: AES, RSA and ECC. In order to achieve this goal, we first find secret dependent instructions/data with dynamic taint analysis, introduce cache evictions inside code routines, record cache traces and then verify the existing dependence with respect to the secret. Our results show that, despite the efforts of cryptography library designers in reaction to the multitude of recently published LLC cache attacks 
several popular libraries are still vulnerable.
\medskip
\noindent
{\bf Our Contributions.}

\begin{packed_itemize}
\item This paper presents a proactive tool to analyze leakage behavior of security critical code. Unlike other approaches, our tool can be used to find real exploitable leakage in the design of cryptographic implementations.
\item The tool identifies secret dependent data, obtains cache traces obtained from the code execution on the microarchitecture and uses the generic Mutual Information Analysis (MIA) as a metric to determine the existence of cache leakage.
\item The detection technique is agnostic to the implementation, i.e. the testing code can be run across all target platforms without having to redesign it, yet pinpoints parts of code that cause found leakages.
\item We perform the first big scale analysis of popular cryptographic libraries on three of the most commonly used primitives: AES, RSA and ECC.
\item We demonstrate that several cache leakages are still present in up-to-date cryptographic libraries (i.e. 50\% still leak information) and need to be fixed.
\end{packed_itemize}

\section{Preliminaries}
The proposed tool employs techniques from  modern cache attacks to monitor cache activity and measures information leakage using Mutual Information.

\subsection{LLC Attacks}
LLC cache attacks are one of the most dangerous side channel attacks since they do not rely on physical proximity. Furthermore, these attacks do not require root privileges and can be executed in user mode. There are two main classes of LLC attacks:

\smallskip
\noindent
{\bf Flush and Reload Attack}
The \texttt{Flush and Reload} attack was first introduced in Gullasch et al.~\cite{Gullasch:2011:CGB:2006077.2006784} but acquired its name in Yarom and Falkner~\cite{184415}, who were able to extract RSA cryptographic keys across VMs situated in different cores. This attack was later shown to be successful also breaking symmetric key algorithms~\cite{waitaminute,191010}, TLS sessions~\cite{Irazoqui:2015:LSB:2714576.2714625} or the isolation in PaaS clouds~\cite{Zhang:2014:CSA:2660267.2660356} and smartphone devices~\cite{DBLP:conf/uss/LippGSMM16,Zhang:2016:RFS:2976749.2978360}. In particular, the attack assumes the following;

\begin{packed_itemize}	
\item The attacker and the victim are executing processes in the same CPU but in different cores. 
\item Attacker and victim share read-only memory pages. 
\end{packed_itemize}

The \texttt{Flush and Reload} attack can only succeed when attacking memory blocks shared with the potential victim. These usually imply static code and global variables, but never dynamically allocated variables.

\smallskip
\noindent
{\bf Prime and Probe Attack}
The \texttt{Prime and Probe} attack was first introduced for the L1 cache by Osvik et al.~\cite{Osvik:2006:CAC:2117739.2117741} and was later utilized by Ristenpart et al.~\cite{Ristenpart_hey} and Zhang et al.~\cite{Zhang:2012:CSC:2382196.2382230} to recover, in IaaS clouds, keystrokes and El Gamal decryption keys respectively. Recently the attack was expanded to attack the LLC by Liu et al.~\cite{lastlevel} and Irazoqui et al.~\cite{sca}, and has been successfully applied in commercial IaaS clouds~\cite{inci2016cache}, as Javascript executions~\cite{DBLP:journals/corr/OrenKSK15} and as smartphone applications~\cite{DBLP:conf/uss/LippGSMM16}. The attack presents advantages and disadvantages over \texttt{Flush and Reload}:

\begin{packed_itemize}
\item \texttt{Prime and Probe} does not require any memory sharing between victim and attacker.
\item \texttt{Prime and Probe} can increase the attack vector to dynamically allocated variables.
\item \texttt{Prime and Probe} implies some reverse engineering to know which sets in the cache the attacker is using, while \texttt{Flush and Reload} does not.
\item \texttt{Prime and Probe} is noisier than \texttt{Flush and Reload}.
\end{packed_itemize}

Taking these two attacks into account, we explore the cryptographic libraries looking for leakages exploitable by \emph{either} attack. Thus, we examine both statically and dynamically allocated memory in our analysis.

\subsection{Mutual Information Analysis}
	The metric that we use to quantify leakage is Mutual Information (MI), that is given by the following equation:
	\begin{equation*} 
	I(X;Y) = H(X) - H(X|Y) = H(X) + H(Y) - H(X, Y)
	\end{equation*}
	Where $H(X)$ is the entropy of random variable $X$ and $H(X|Y)$ is the entropy of the random variable $X$ given the knowledge of the random variable $Y$. In a way $I(X; Y)$ gives us how related variables $X$ and $Y$ are. Note that if $X$ and $Y$ are independent, $I(X;Y)=0$ since $H(X|Y)=H(X)$. In contrast, if $X$ and $Y$ are fully dependent we obtain maximum MI, i.e., $I(X;Y)=H(X)$, since $H(X|Y)=0$ ($Y$ fully determines $X$).

	MI has been used in prior work for side channel attacks and leakage quantification. Gierlichs et al.~\cite{Gierlichs2008} utilize MI as a side channel distinguisher to mount differential side channel attacks. Standaert et al.~\cite{Standaert2009} utilized MI as a leakage quantifier to evaluate side channel attack security. Prouff et al.~\cite{Prouff2009} further expanded on the limitations and strengths of MI as a side channel attack metric. Recently Zhang and Lee~\cite{Zhang:2014:NMC} used MI to measure the security level of several cache architectures which they modeled as finite state machines. 
\section{Methodology}
Our goal is to determine whether specific memory addresses being used inside a cryptographic routine reveal information based on their cache accesses. 

\subsection{Main Idea}
Our approach involves the monitorization of the cache usage for all the memory addresses considered susceptible of carrying information. In order to achieve this goal we utilize common known techniques from cache attacks, like the usage of cache flushing or cycle counter instructions.

We illustrate our approach on a toy example shown in Figure~\ref{code:example}. The \texttt{Hello} code snippet simply returns different messages depending on whether the caller is a female or a male. We assume that the designer of this simple code does not want a potential attacker to know the gender of the caller. Assume that the code designers would like to know whether they did a good job, i.e., whether the cache traces do not reveal the gender of the user. Here, it is easy to observe that there is a $gender$ dependent branch that could reveal whether the user is a male or female. We call $B1$ and $B2$ the two possible outputs of the branch.
\begin{figure}[thp]
\begin{CenteredBox}
\begin{lstlisting}[numbers=left,language=c,basicstyle=\small]
void Hello(char *gender){
  if (gender=="Male"){
    male=1;
  }
  else if (gender=="Female"){
    female=1;
  }
}
\end{lstlisting}
\end{CenteredBox}
\caption{Vulnerable code snippet example}
\label{code:example}
\end{figure}
To test the sanity of the code, we insert forced evictions at the beginning of the code routine for those memory references susceptible of leaking information about the gender, and time the re-access after its execution, as shown in Figure~\ref{code:example2}. Then we re-execute the code routine for several input values, retrieving the cache traces for all those potentially leaky references.
%
%

\lstset{moredelim=[is][\color{red}\bfseries]{[*}{*]},numbersep=5pt}
\lstset{moredelim=[is][\color{green}]{/*}{*/},numbersep=5pt}

\begin{figure}[thp]
\begin{CenteredBox}
  \begin{lstlisting}[numbers=left,basicstyle=\small]
[*evict(B1, B2);*]
Hello(&gender);
[*time(access(B1, B2));*]
  \end{lstlisting}
\end{CenteredBox}
\caption{Code snippet wrapped in cache tracer\vspace{-3ex}}
\label{code:example2}
\end{figure}

\begin{figure}[ht!]
  \centering
	\includegraphics[width=0.8\linewidth]{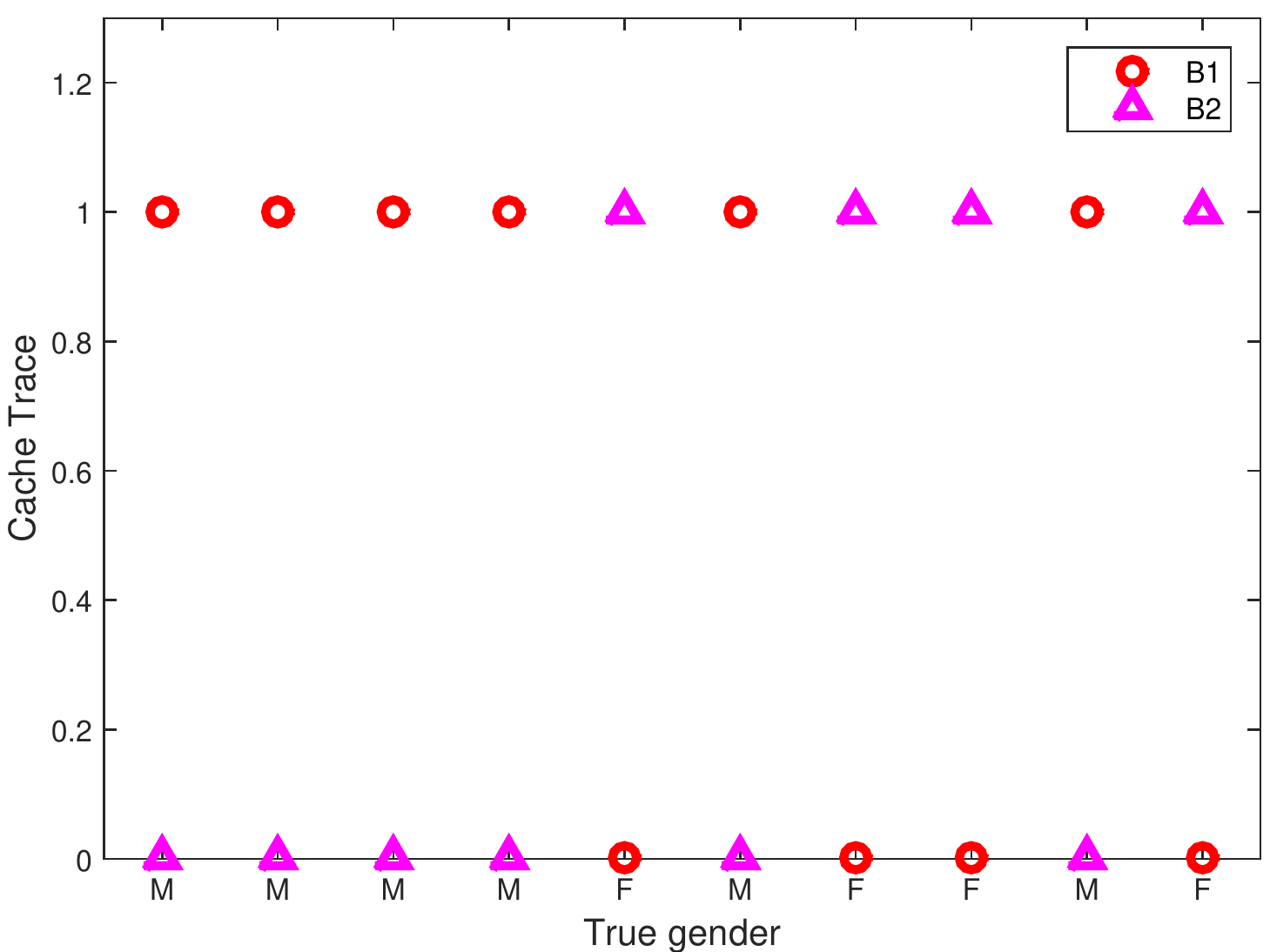}
	\caption{Results for cache traces obtained from invocations of Hello for varying inputs. The x-axis shows the actual gender whereas the y-axis represents cache hits/misses (1 indicates hit, 0 indicates miss). The cache trace has full correlation with $B1$ and $B2$ traces.\vspace{-3ex}}.
	\label{fig:hfm}
\end{figure}

The cache traces for this simple example are shown in Figure~\ref{fig:hfm}. The $x$-axis shows the actual gender while the $y$-axis represents accesses either to the cache as a 1 or to memory as 0. The branch outputs are represented with different colors. Observe that the cache traces match perfectly the gender that the code designer is trying to hide. Thus, this code snippet would be classified as vulnerable. This is because $B1$ and $B2$ were located in different cache lines. However, due to their proximity, it could have happened that both fall in the same cache line, in which case the code would be classified as non-leaky.

Although identifying leakages in the toy example did not seem overly challenging, as the code starts getting more and more complicated, it becomes increasingly difficult and time consuming both to find secret dependent memory and to detect the leakages using manual code inspection. Particularly difficult to detect are cases in which the code does not seem to leak, but subtle architectural optimizations at the hardware level make the leakage appear~\cite{Irazoqui:2015:LSB:2714576.2714625}. In order to cope with this issues we introduce a tool that automatically finds secret dependent memory thanks to the properties of dynamic taint analysis and that automates the analysis of cache leakages through the usage of statistical properties like Mutual Information for number of different inputs. We proceed to explain the metric and the methodology to identify secret dependent memory locations and to uncover hidden cache leakages.

\subsection{Our approach: MI based Leakage Detection}


Informally we can summarize our technique as follows: We first identify, thanks to dynamic taint analysis, potentially leaky pieces in the target code that may compromise the secret. For instance, \emph{variables} and \emph{branch instructions} that can reveal information to the attacker. Then we insert forced cache hierarchy evictions on those secret related memory addresses prior to the execution of the targeted piece of code, and accesses after it. These accesses determine a clear cache fingerprint of the potentially leaky memory addresses. Finally, we correlate the observed cache trace with the secret to be extracted by the attacker using Mutual Information. We summarize the proposed approach in four steps:

\begin{enumerate}

\item Identify, through a dynamic taint analysis, {\em suspect instructions and variables} that might reveal information related to a secret in the targeted code.

\item Insert forced cache hierarchy evictions just before the vulnerable targeted code, and accesses after it, for all the variables/instructions susceptible of leaking secret information.

\item Execute the targeted code under different secret values, and record the cache traces for every possible vulnerable variable.

\item Correlate, with MI, the observed cache traces to the secret value to determine leakage existence.
\end{enumerate}

\subsection{The Detection Algorithm} 

We formalize the detection process as shown in Algorithm~\ref{algo:main}. The input is the fragment of code that we want to analyze. 
First, the first step involves a dynamic taint analysis to retrieve the potential variables/instructions that are dependent on the secret the code routine is hiding. For each of them we generate access traces obtained through the execution of the code for a large number of secret values. We then compute the mutual information between the secret value used and access trace obtained through during an execution of the code. We characterize the code as to leak information if the average mutual information is above a predefined baseline $threshold$. We should note that two critical steps require code inspection: identification of the suspect variables/instructions and how the secret will be processed during MIA. Splitting the key requires understanding of the cryptographic algorithm. To improve detection rates we split the key in a way that mimics key is processing in a particular library (more details in Section~\ref{sec:primitives}).

	\begin{algorithm}
		\SetKwInOut{Input}{Input}
		\SetKwInOut{Output}{Output}
		
		\underline{function DoesItLeak} $(C)$\;
		\Input{Code $C$}
		\Output{True: the code leaks; \\ False: the code does not leak}
    $L =$ ListSecretTaintedMemory($C$)\;
		\For{{\bf each} $vi$ {\bf in} $L$}{
			trace $= [~]$, M $= [~]$\;
			\For{{\bf each} $key=\langle k_0,k_1,\ldots, k_{\ell}\rangle$ value}{
				{\bf evict}(vi)\;
				{\bf execute}($C$, $key$)\;
				trace.append({\bf time}($vi$))\;
				M.append(MI(trace, $\langle k_0,k_1,\ldots, k_{\ell}\rangle$)\;
			}
			\If{$M$.average$>threshold$}{
				return True\;
			}
		}
		return False\;
		\caption{MI based leakage test}
		\label{algo:main}
	\end{algorithm}

\subsection{Secret Dependent Memory Detection using Dynamic Taint Analysis}
In this work, we use dynamic taint analysis to effectively identify suspect instructions and variables.
In the past decade, dynamic taint analysis has been widely explored in many research areas, 
including malware analysis, vulnerability detection, program debugging and data leak protection~\cite{Zhu:2011:TPS:1945023.1945039, TaintPipe}.
Dynamic taint analysis is a powerful approach to track the propagation of sensitive program data such as user inputs.

To detect key dependent memory for crypto libraries, our dynamic taint analysis retrieves two kinds of information. 
First, we collect all branch instructions affected by the taint inputs.
Second, we collect all memory object information that is accessed with a taint value as index.

We illustrate our approach on a toy example shown in Figure~\ref{code:taintexample}. 
The \verb+encrypt+ function takes user input and encryption keys, and returns \verb+out+ including encrypted data.
The \verb+encrypt+ function checks \verb+key+ and decides which variable is used for encryption and invokes \verb+compute+ function to compute output.
In the \verb+encrypt+ function, variable \verb+key+ is marked as taint inputs.
Then we execute the program using our taint analysis engine. 
The engine can easily identify that Line 8, variable index (Line 13) and variable A (Line 14) are our desired information. This does not mean that all three memory addresses will leak information; in fact, the variable index will not leak information as it is used in both cases. Our taint analysis outputs secret dependent data that might not leak information, and thus has to be verified  with cache trace analysis.

%
\begin{figure}[thp]
\begin{CenteredBox}
\begin{lstlisting}[numbers=left,language=c,basicstyle=\small]
static int A[] = {0x1, 0x2, 0x3, 0x4};
static int B = 0x5;

void encrypt(int *in, int *out, int *k)
{
int index
  for(int i = 0; i < 64; ++i) {
   if(k[i]>=32) {
      index = 0;
      compute(in, out, B);
    }
    else {
      index = k[i] % 4;
      compute(in, out, A[index]);
    }
  }
}

void main()
{
  int key[64], *in, *out;
  get_input(&key, in, out);
  set_taint(&key, 64, "key");

  encrypt(in, out, key);
}
\end{lstlisting}
\end{CenteredBox}
\caption{Taint analysis example}
\label{code:taintexample}
\end{figure}

We employ a symbolic execution engine KLEE~\cite{Cadar:2008:KUA:1855741.1855756} as our taint analysis engine. 
The target program is compiled into a LLVM intermediate representation, and then executed by the taint analysis engine to collect suspect instructions and variables.

\subsection{Collecting the Cache Traces}
We collect the cache usage for those memory locations that our dyncamic taint analysis outputs during different executions for varying secret inputs of the targeted code. In order to mimic the attacker ability to insert evictions, and due to its high resolution, we utilize the common cache flushing and timing approach typically utilized in LLC attacks. In short, we insert forced flushes (with the \texttt{clflush} instruction) for those variables that have a dependency with the key before the targeted code, and timed accesses after it. These timing measurements are later converted to 0s and 1s, compared to a pre-defined memory access threshold, indicating whether these memory addresses were retrieved from the cache or from the memory. These cache traces are later correlated using MI statistical properties to derive dependencies with the secret to hide.
\subsection{Dealing with noise} 
\begin{figure}[ht]
	\centering
	\includegraphics[width=\linewidth]{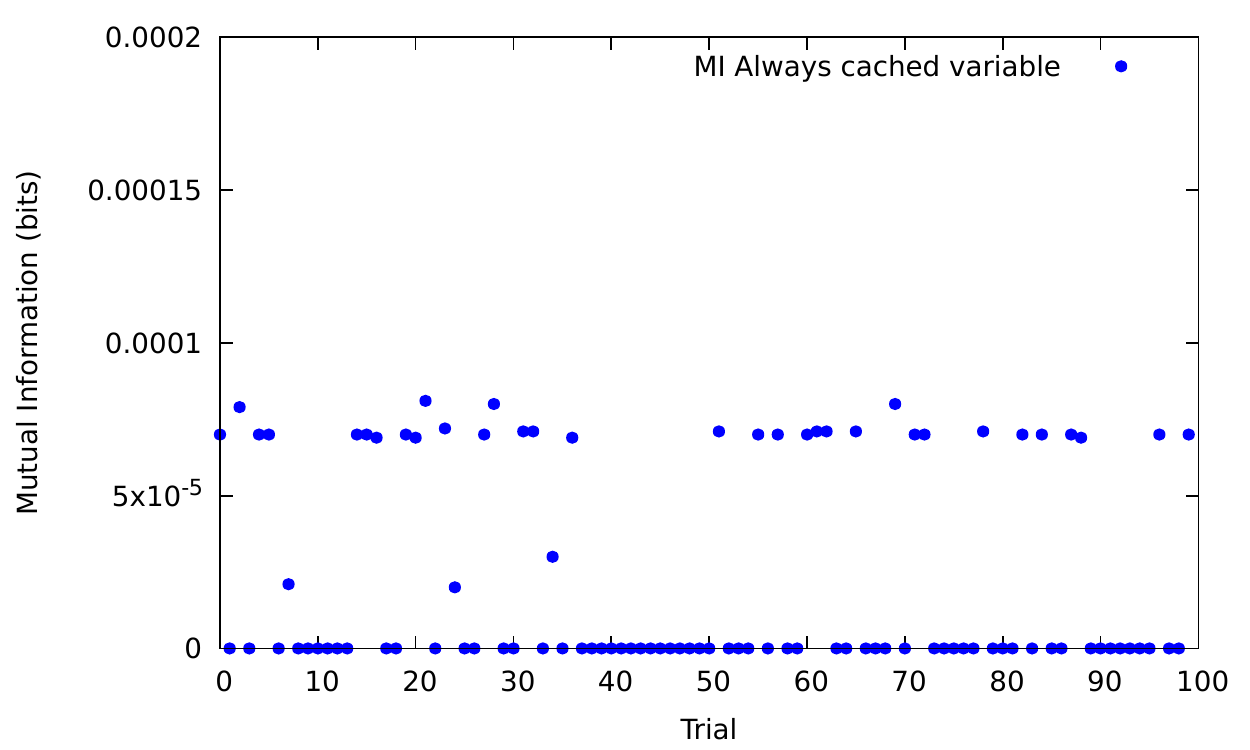}
	\caption{100 MI measurements each composed of $10^5$ time accesses to a cached variable. Our noise threshold is set to $\mu_{noise}+3\sigma$}
	\label{fig:nois}
\end{figure}

Although our methodology works best in an environment with minimal unintended noise coming from co-resident processes, our measurements will still suffer from the omni-present microarchitectural/ OS noise. In that sense, the $threshold$ (see Algorithm~\ref{algo:main}) for which an implementation will be considered to be leaky has to be established taking into account these unintended noise sources.
Aiming at correctly categorizing the leakage, we perform a set of identical measurements to those carried out for the cryptographic primitives to an \emph{always used} variable under microarchitectural noise. Note that the variable is always cached and thus it should not present any leakage. Our goal is to see how minimum microarchitectural noise affects to the measurements that will later be taken for the cryptographic primitives. The results of 100 MI calculations under minimal microarchitectural/OS noise on datasets containing $10^5$ cache access times can be observed in Figure~\ref{fig:nois}. Our noise threshold will follow the famous \emph{three-sigma rule of thumb}~\cite{Pukelsheim94}, although one can use any statistical method that fits best to the approach taken. 
Finally we note the threshold we set via experimentation is rather conservative. This stems from the fact that cache access timing measurements are susceptible to have false negatives, while memory accesses timing measurements do not show false negatives. Therefore, microarchitectural noise tends to affect timing measurements by {\em increasing the measured time}, but rarely reducing it.

%

\section{Evaluated Crypto Primitives \label{sec:primitives}}

We restrict our evaluation to three of the most widely used crypto primitives: AES, RSA and ECC. Note that this methodology can be extended to any cryptographic software code. However, for brevity we only focus on these three primitives. Note that each cryptographic primitive has its own leakage behavior. Different implementations of the same cryptographic primitive may also result in very different leakage behavior. Our analysis excludes the quantization of the number of measurements needed by an attacker to exploit the leakage. Indeed, we believe code designers should aim at designing leakage free code without taking into account the attacker exploitation effort.
Our analysis was performed on both Intel Core i7-6700K Skylake and  Intel Core i5-650 Nehalem processors. However, the leakage we found does not seem to be architecture dependent, except those for which the cache line size affected the leakage. Since the cache line size is 64 bytes in the three most widely used processors, i.e., Intel, ARM and AMD, our analysis should detect the same leakages in other processors. 

\subsection{AES}
Across the cryptographic libraries evaluated, we found three different AES implementations. The AES-NI implementation will be excluded from the analysis, since it is a pure hardware based implementation. We are left with two techniques to implement AES:
\begin{itemize}

\item{\bf T-table implementation:} The algorithm utilizes 256 entry tables, each entry holding a 32 bit value.

\begin{figure}[t!]
	\centering
	\includegraphics[width=0.7\linewidth]{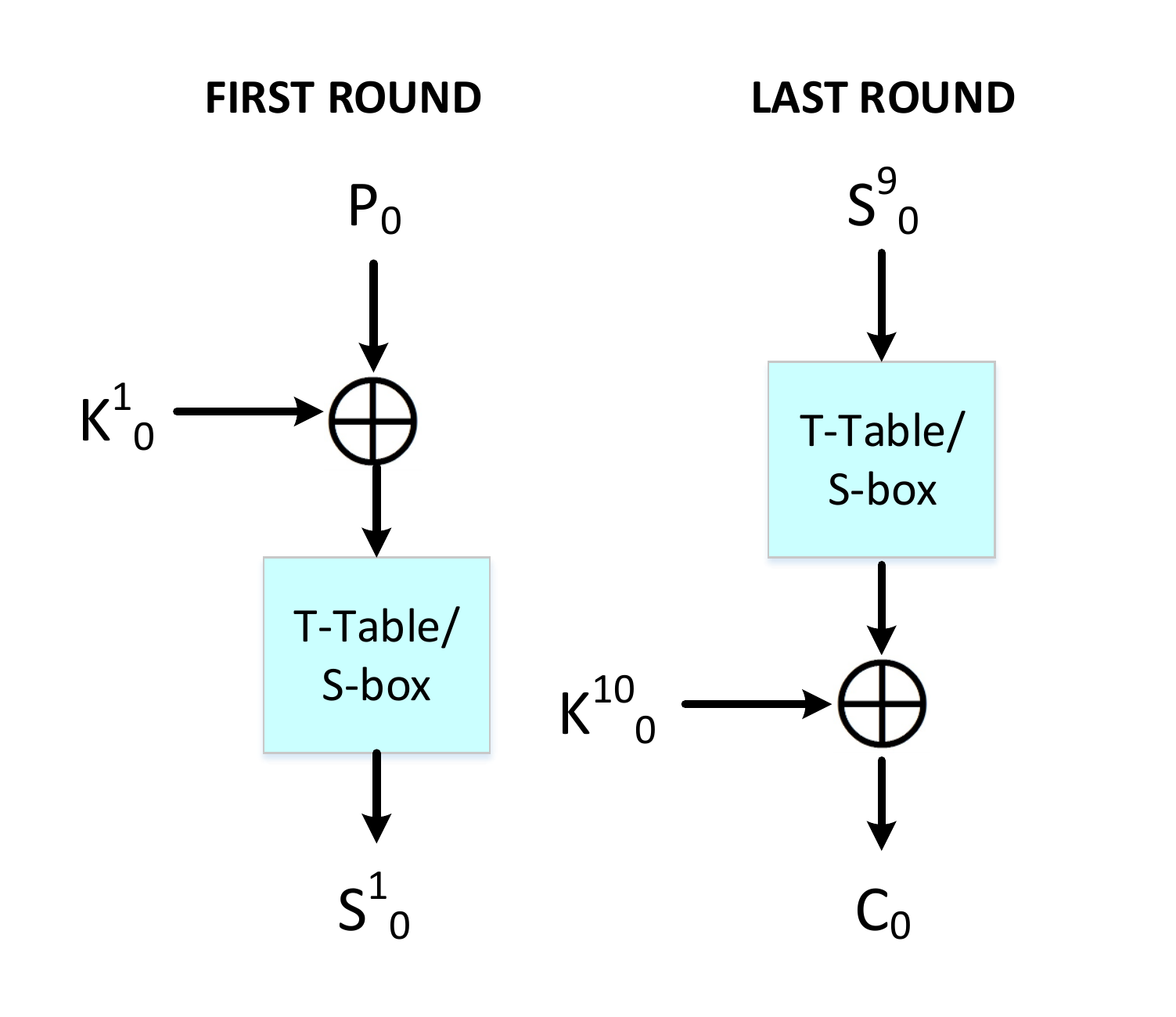}
	\caption{First and Last round of an AES encryption. In both cases, the entry to the T-table/S-box is influenced by the key and the plaintext/ciphertext}
	\label{fig:first_last}
\end{figure}
	
\item{\bf S-box implementation:} The algorithm utilizes a single 256 entry table for the whole encryption, each entry holding an 8-bit value~\cite{Daemen:2002:DR:560131}.
\end{itemize}

Some implementations also might use a combination of both approaches, e.g., T-tables in the first rounds and S-box in the last. Note that these represent only the implementations we encountered across the libraries we evaluated, but are not the only  implementation options of AES. For instance, AES routines may utilize S-box/ T-table free implementations, which compute the inverse in Galois field mode without pre-computed tables.

\medskip
\noindent
{\bf Customizing for AES Leakage Detection:}
For AES, the approach we follow is: 
\begin{packed_itemize}
\item the plaintext is fixed; 
\item for different key byte values we record different cache traces of our potentially leaky variables; and
\item the traces and the key bytes are correlated
\end{packed_itemize}
However, note that we can also fix the key and randomize the plaintext input byte, since in the first round the T-table entry access is influenced by both the plaintext and the key bytes (see Figure~\ref{fig:first_last}). Similarly, we can apply the same concept in the last round and compute the MI of our cache traces with the last round key bytes or with the ciphertext bytes, since again the last round T-table access is influenced by both the ciphertext and last round key bytes (Figure~\ref{fig:first_last}). This might be convenient in cases where the last round is handled differently than the first 9 rounds and leaks more information than the first. In this work, we compute the MI test in both the first and last rounds, and select the one that leaks more information.

We restrict our analysis to $10^5$ and $10^7$ encryptions. This restriction allows us to test AES leakages in at most a few hours. Furthermore, cache side channel attacks performed against AES have succeeded with fewer number of traces than those we require in our analysis, c.f.~\cite{waitaminute}.
\subsection{RSA}
The second cryptographic primitive that we evaluate is RSA, the most widely used public key algorithm. RSA can be implemented in several ways, including by using the Chinese Remainder Theorem, with and without ciphertext blinding or exponent blinding. The only implementation that would completely stop cache and other side channel attacks from recovering meaningful information is exponent blinding, since a different exponent is used for every decryption. However, we have not seen any cryptographic library using exponent blinding. RSA leakages can come from any key dependent execution or memory usage during the modular exponentiation. 

\medskip
\noindent
{\bf Customizing for RSA Leakage Detection:}
Across all the libraries evaluated in this work, we found that there are two main (and distinct) ways in which crypto libraries implement modular exponentiation in practice:
\begin{packed_itemize}
	\item{\bf Bitwise modular exponentiation}: The algorithm processes the key and executes all the squaring and multiplication operations bit by bit. This means, in our detection algorithm we can split the key into bits and then correlate with MIA with traces collected over the entire RSA encryption.
	\item{\bf Windowed modular exponentiation}: The algorithm first precomputes a set of values ($b^0,b^1,$\ldots$,b^{2^w-1}$ for fixed window, $b^{2^{w-1}},b^{2^{w-1}} -1,$\ldots$,b^{2^w-1}$ for sliding window) that are stored in a table. Then, the key is processed in chunks of at most $w$ bits. For each chunk, at most $w$ squares are executed together with a multiplication with the appropriate window value in the precomputed table, see Algorithm. Our detection algorithm therefore, is customized to split the key into fixed/sliding windows which are then correlated via MIA with access traces collected over the entire windowed RSA encryption.

\end{packed_itemize}
Our analysis consists of observing leakage at the decryption process for 100 different 2048 bit decryption keys. Note that, unlike AES, RSA leakages are visible with much fewer encryptions since the algorithm leaks serially bit by bit. AES on the contrary needs aggregated measurements to derive the table entries utilized.

\subsection{ECC}
The last cryptographic primitive we analyzed is Elliptic Curve Cryptography (ECC). We utilize Elliptic Curve Diffie Hellman (ECDH) as opposed to Elliptic Curve Digital Signature Algorithm (ECDSA) because in the latter the secret operation is performed on an ephemeral key, which limits our analysis capabilities. Note that this does not imply that the leakage observed for ECDH will not be present in ECDSA, since both algorithms use the same implementation techniques in all the libraries analyzed. As before, we utilize 100 different 256-bit private keys.

\medskip
\noindent
{\bf Customizing for ECC Leakage Detection:}
The scalar point multiplication implementations we found are very similar to those observed for modular exponentiation:
\begin{packed_itemize}
	\item{\bf Bitwise Double and Add}: The algorithm processes the secret scalar bit by bit and executes double and add operations accordingly. In the detection algorithm the key is split into bits and then correlated with access traces obtained over a scalar point multiplication computation using MIA.
	
	\item{\bf Windowed algorithm}: Similar to the case of RSA, the algorithm  first pre-computes the set of values ($P,2P,..$\ldots$,(2^w-1)P$ for fixed window, $(2^{w-1})P,(2^{w-1}+1)P, $\ldots$,(2^{w}-1)P$ for sliding window and $(P,3P, $\ldots$,(2^{w-1}-1)P$ for wNAF) that are stored in a table, being $w$ the window size. Then, the algorithm processes the secret scalar in chunks of at most $w$ bits. For each chunk, at most $w$ doublings are executed and an addition with the appropriate value of the window in the precomputed table. The detection algorithm splits the key into windows and is correlated with access traces obtained over a scalar point multiplication using MIA.
	
\end{packed_itemize}
	
\subsection{Cryptographic libraries evaluated}
Our evaluation is restricted to 8 popular up-to-date (at the time of the leakage evaluation) cryptographic libraries, represented in Table~\ref{tab:2}. The analysis can be extended to other cryptographic libraries with mild effort. 

\begin{table}[]
	\centering
	\resizebox{!}{1.5cm}{%

  \begin{tabular}{|l|r|}\hline
	\bf Cryptographic Library & \bf Version \\
	\hline
  OpenSSL & 1.0.1-t\\
	wolfSSL & 3.9.0\\
	Intel Integrated Performance Primitives (IPP) & 9.0.3\\
	Bouncy Castle & 1.8.1\\
	LibreSSL & 2.4.2\\
	Mozilla Network Security Services (NSS) & 4.13.0\\
	Libgcrypt (NSS) & 1.7.3\\
	mbedTLS & 2.3.0 \\\hline
\end{tabular}}
	\caption{Cryptographic libraries evaluated\vspace{-3ex}}
	\label{tab:2}
\end{table}
\section{Results for AES}

\begin{figure*}[h!]
    \centering
		\subfloat[]{
                \includegraphics[width=0.44\linewidth, height = 0.2\textheight, keepaspectratio=true]{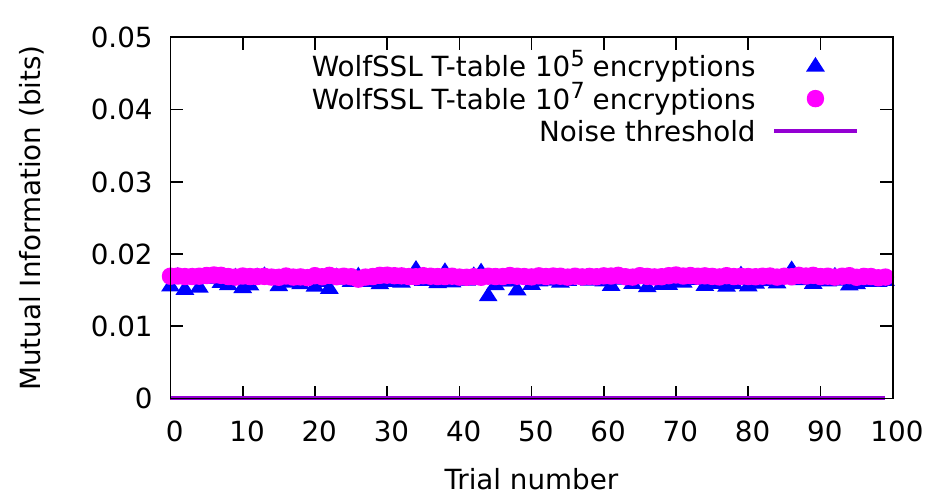}
\label{fig:waes1}}
	  \subfloat[]{
                \includegraphics[width=0.44\linewidth, height = 0.2\textheight, keepaspectratio=true]{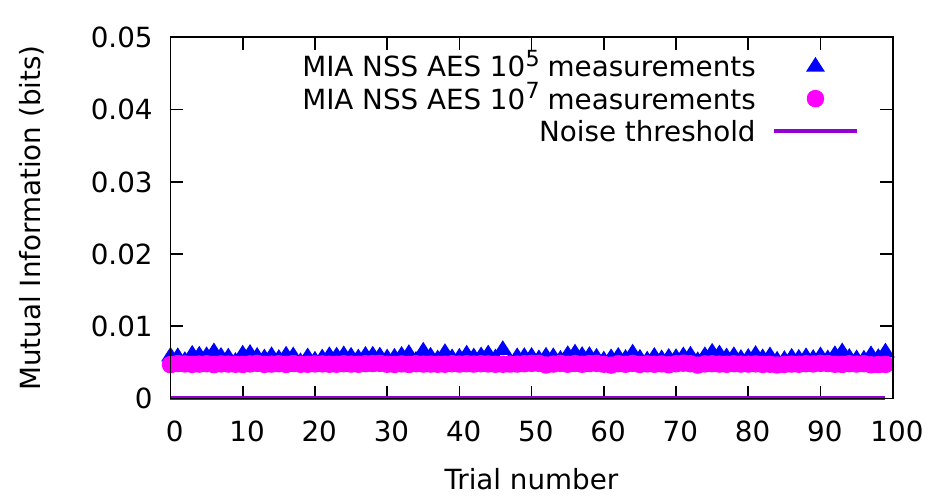}
\label{fig:waes2}}
\caption{AES MI leakage in a) wolfSSL and b) NSS. The leakage is observable at $10^5$ encryptions in both cases.}
\end{figure*}

\begin{figure*}[h!]
  \centering

	\subfloat[]{
							  \centering
                \includegraphics[width=0.44\linewidth, height = 0.2\textheight, keepaspectratio=true]{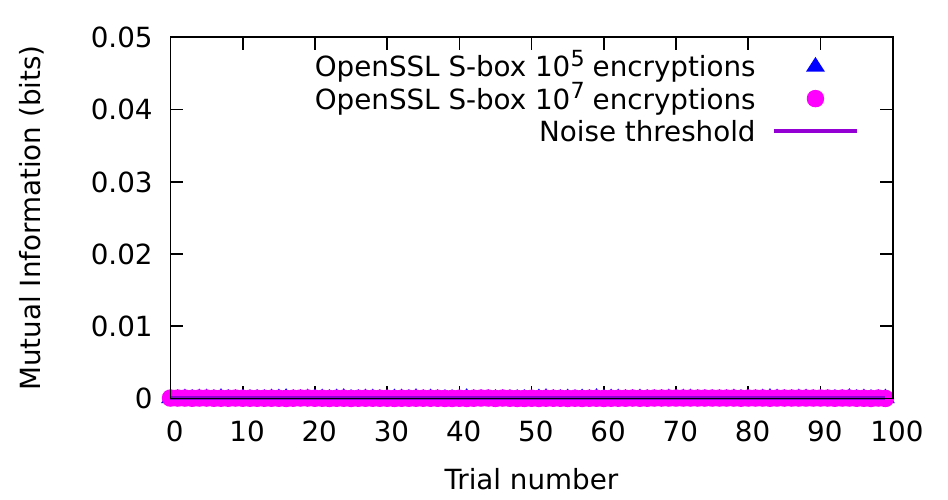}
								
						\label{fig:oaes1}}
	\subfloat[]{					
							  \centering
                \includegraphics[width=0.44\linewidth, height = 0.2\textheight, keepaspectratio=true]{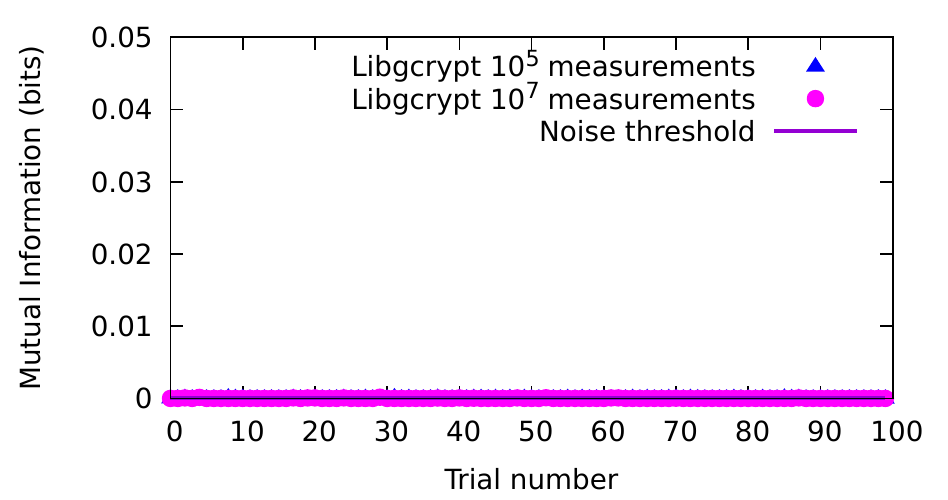}

						\label{fig:oaes2}}
						
\caption{AES S-box implementation MI leakage in a) OpenSSL and b) Libgcrypt. None of them leak information.}
\label{fig:oaes}
\end{figure*}

In this section we present the results obtained for AES implementations for different libraries. Due to space constraints, we do not include all the results analyzed, but rather examples of every kind of implementation. The complete results are summarized in Section~\ref{sec:sum}. 

\subsection{T-table based AES (wolfSSL, mbedTLS, NSS \& Bouncy Castle)}

wolfSSL, mbedTLS, NSS and Bouncy Castle utilize T-table based implementations to perform AES encryptions. We only include examples for wolfSSL and NSS. Note that the results obtained for the other libraries are similar to the ones obtained for wolfSSL and NSS shown in Figure~\ref{fig:waes1} and Figure~\ref{fig:waes2}, respectively. For both cases, the T-tables were classified as key dependent accesses and leaked information. Observe that in both cases the leakage is already observable at $10^5$ encryptions, and that wolfSSL seems to leak more information than NSS. This is because wolfSSL uses a different T-table for the last round, while NSS re-uses the first 4 T-tables in the last round. In consequence, the probabilities are more distinguishable in the case of wolfSSL. In both cases the leakage is above our noise threshold thus leading to key recovery attacks after a small number of encryptions.

\subsection{S-Box based AES (OpenSSL, LibreSSL and libgcrypt)}
In this case we analyze those implementations that utilize a single 256 entry S-Box in the AES encryption. As before, we only show the results obtained for two of them, since the results obtained for the rest are very similar. In this case, the results for OpenSSL and Libgcrypt are included in Figure~\ref{fig:oaes1} and Figure~\ref{fig:oaes2}, respectively. In both cases, the S-boxes accesses were flagged as key dependent, but the leakage results were not higher than our noise threshold, and as such are categorized as non-leaky implementations. In the case of OpenSSL, we observe that the S-Box entries are prefetched to the cache prior to the encryption. Libgcrypt utilizes the approach described in~\cite{Hamburg2009}, i.e., they use vector permute instructions not only to speed up the AES encryption but further to protect against cache attacks.

\section{Results for RSA}

\begin{figure*}
  \centering
	\subfloat[]{
							  \centering
                \includegraphics[width=0.44\linewidth, height = 0.2\textheight, keepaspectratio=true]{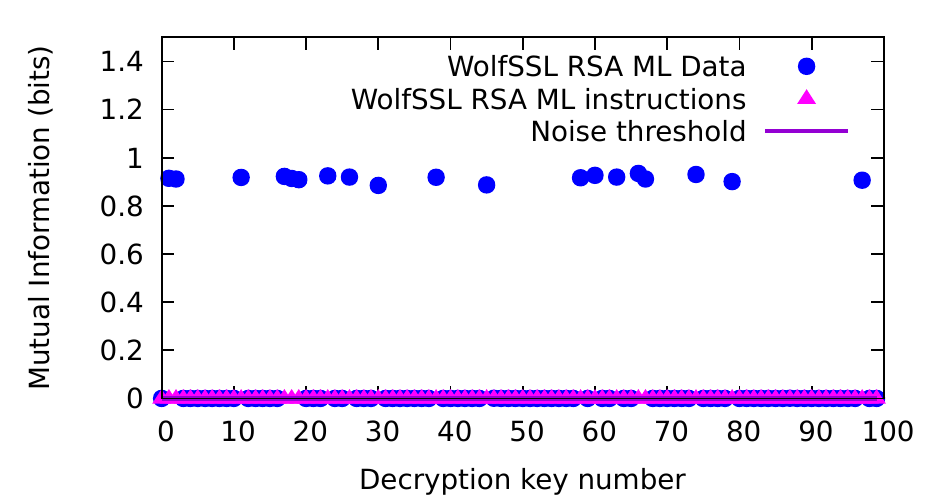}

						\label{fig:montrsa}}
	\subfloat[]{					
							  \centering
                \includegraphics[width=0.44\linewidth, height = 0.2\textheight, keepaspectratio=true]{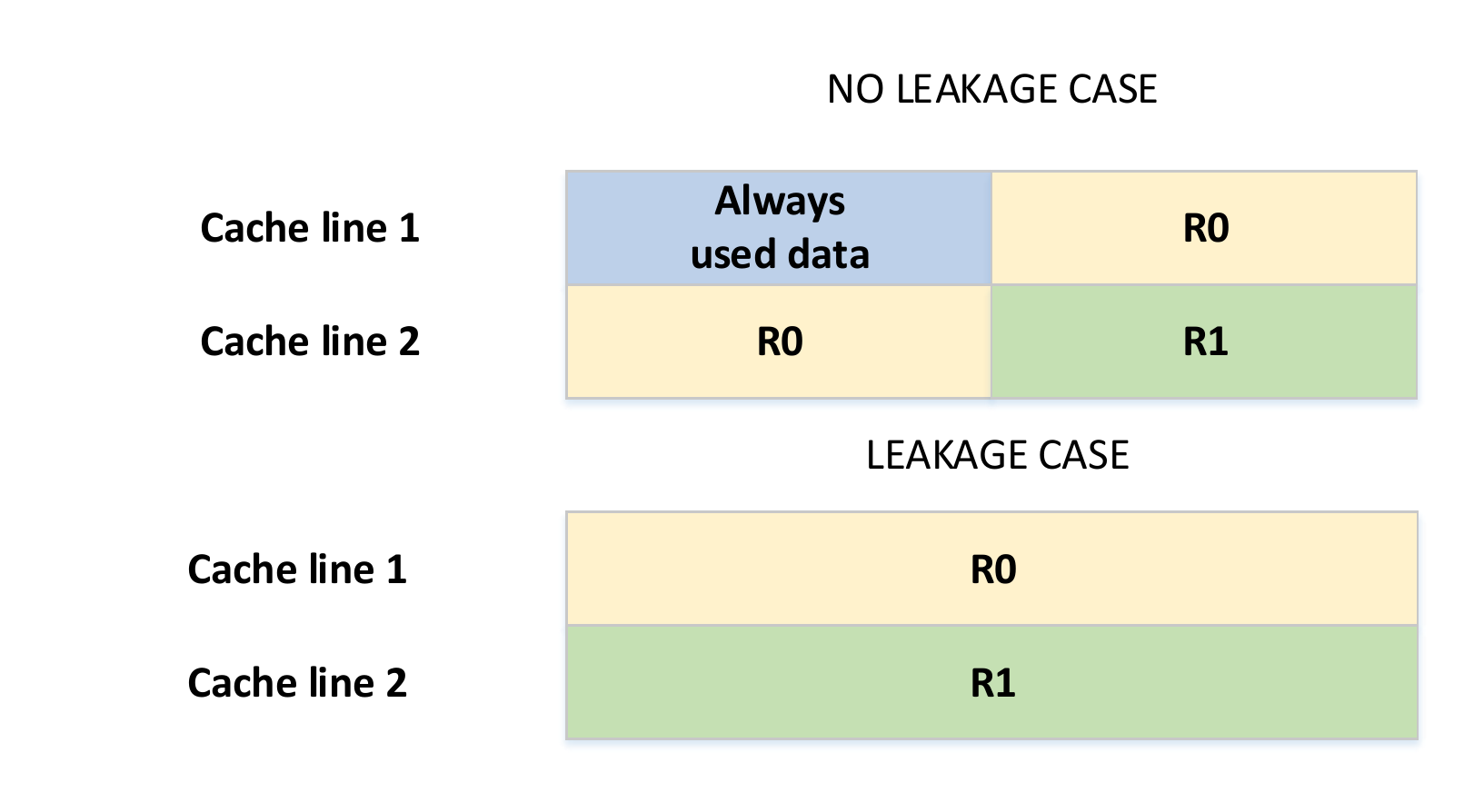}

						\label{fig:wolfexpla}}
						
\caption{Montgomery ladder RSA MI leakage for wolfSSL (a) Instruction leakage is shown in magenta, data leakage shown in blue. Register accesses leak for some keys due to cache alignment issues. b) Explanation for the varying leakage. When R0 is not stored at a cache line boundary, its cache line is always used and we cannot find leakage.}
\label{fig:wolfmont}
\end{figure*}

\begin{figure*}[h!]
  \centering
	\subfloat[]{
									\centering
                \includegraphics[width=0.44\linewidth, keepaspectratio=true]{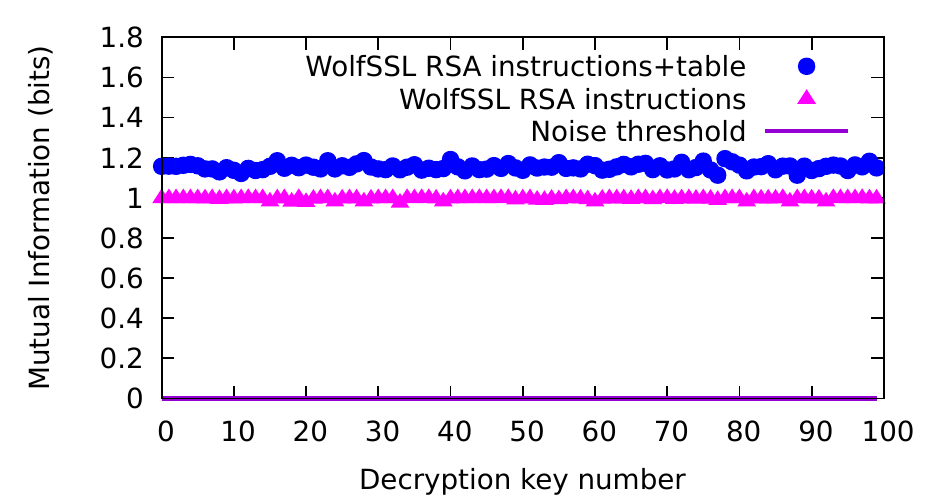}

						\label{fig:polandwolf1}}
	\subfloat[]{					
							  \centering
                \includegraphics[width=0.44\linewidth, keepaspectratio=true]{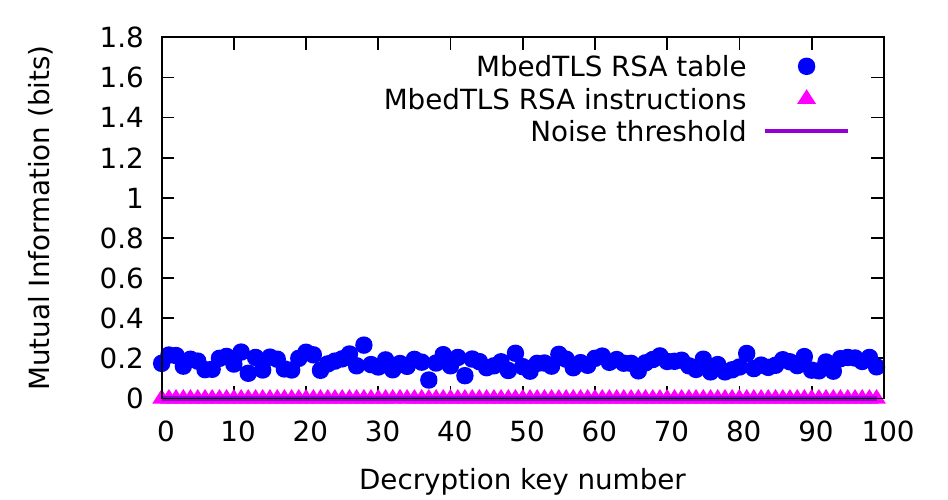}

						\label{fig:polandwolf2}}
						
\caption{Sliding window RSA leakage for a) wolfSSL and b) mbedTLS. Magenta indicates instruction leakage, blue indicates instruction+data leakage. WolfSSL leaks from both possible sources (instructions reveal zeroes between windows, data accesses reveal window values) and mbedTLS only leaks from the precomputed table accesses.}
\label{fig:polandwolf}
\end{figure*}

Similar to the measurements taken for AES, we proceed to test the robustness of the RSA implementations observed across our evaluated libraries. The three main approaches found were:

\subsection{Montgomery Ladder RSA (WolfSSL)}\label{sec:montrsa}

WolfSSL is the only library that utilizes Montgomery ladder to perform modular exponentiations. This implementation attempts to prevent the leakages coming from simple square and multiply approaches. Our results show that, among others, squaring or multiplying instructions and key storing registers are key dependent. The results for these are shown in Figure~\ref{fig:wolfmont}, where instruction leakage is presented in magenta and data leakage coming from the registers in blue. We observe 0 MI for these (or any) instructions involved in the modular exponentiation process leak information. However, in the case of the registers, we see that the MI is 0 for most of the keys, except for some for which it increases to almost 1, the maximum entropy. These measurements cannot be outliers, since even if a noisy measurement is observed, it should not output an almost maximum MI. Thus, after inspecting this particular behavior, we realized that the MI was high \emph{only} when the variable $R_0$ started at a cache line boundary. If it does not, then there is data allocated in the same cache line that is used regardless of the key bit (fetching the $R_0$ cache line always to the cache). This behavior is represented in Figure~\ref{fig:wolfexpla}. Note that the alignment of the register $R_0$ is controlled by the OS, and usually changes from execution to execution. In consequence, a potential attacker only needs to keep iterating until observing a trace for which $R_0$ is aligned with a cache line boundary. Our methodology was able to find this particular cache alignment dependent leakage, which would not be observable without empirical cache traces.

\begin{figure*}[t!]
  \centering
	\subfloat[]{
							  \centering
                \includegraphics[width=0.44\linewidth,  height = 0.2\textheight, keepaspectratio=true]{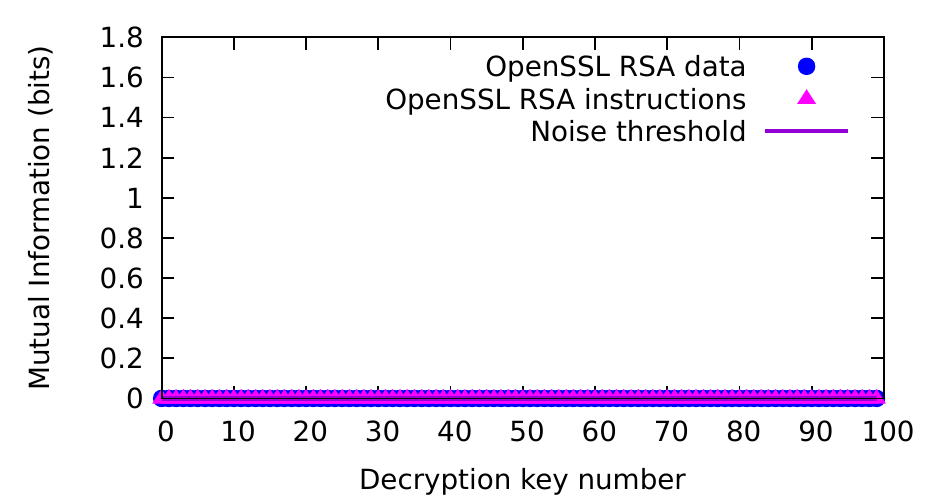}

						\label{fig:opandipp1}}
	\subfloat[]{					
							  \centering
                \includegraphics[width=0.44\linewidth,  height = 0.2\textheight, keepaspectratio=true]{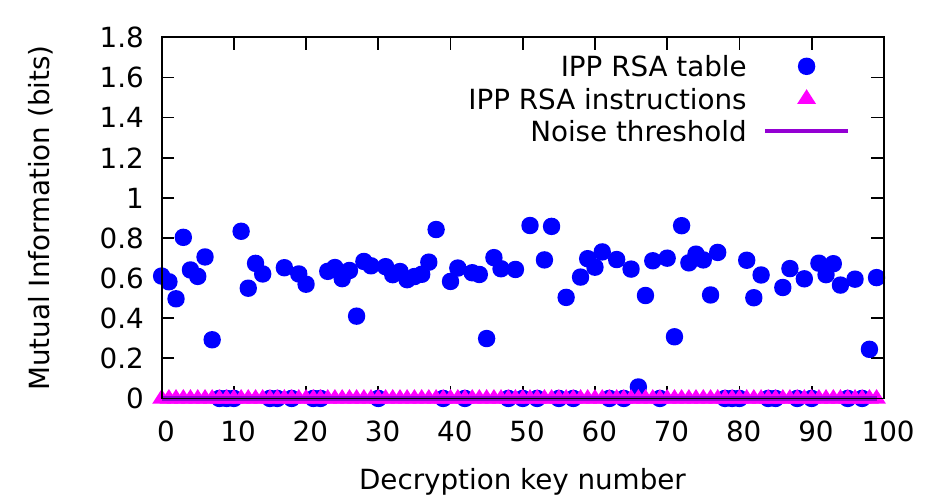}

						\label{fig:opandipp2}}
						
\caption{Leakage for fixed window RSA in a) OpenSSL and b) IPP. Magenta indicates instruction leakage, blue indicates instruction+data leakage. OpenSSL does not leak information (scatter and gather approaches correctly used), while IPP leaks varying amount of leakage for different decryption keys (although scatter and gather is used)}
\label{fig:opandipp}
\end{figure*}

\begin{figure*}[h!]
\centering
	\subfloat[]{
							  \centering
                \includegraphics[width=0.44\linewidth, height = 0.18\textheight, keepaspectratio=true]{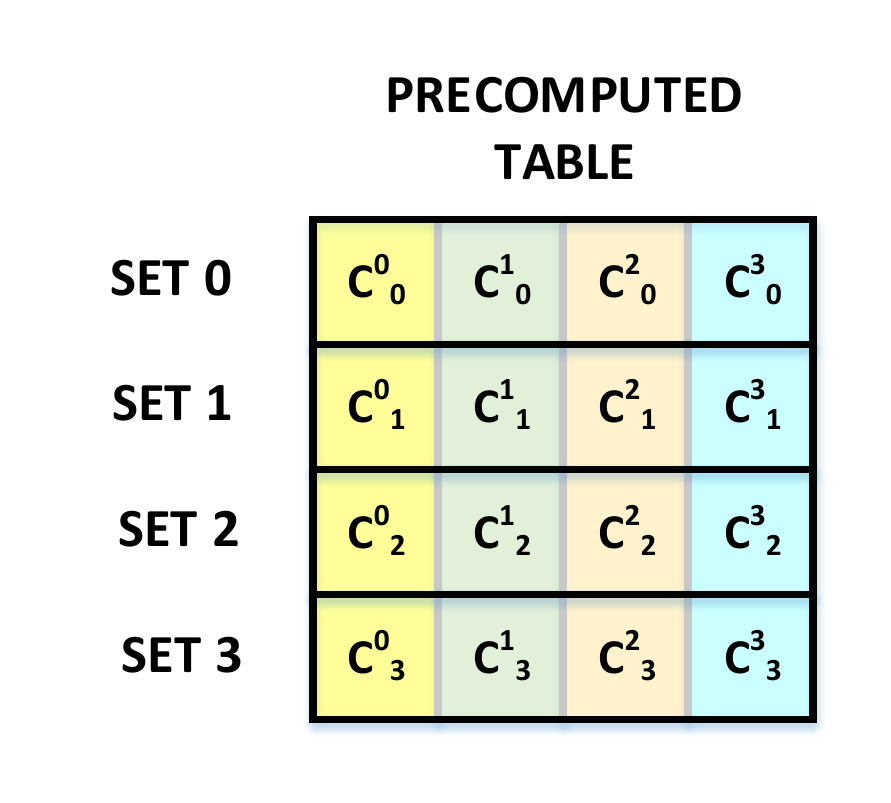}

						\label{fig:aligned}}
	\subfloat[]{					
							  \centering
                \includegraphics[width=0.44\linewidth, height = 0.18\textheight, keepaspectratio=true]{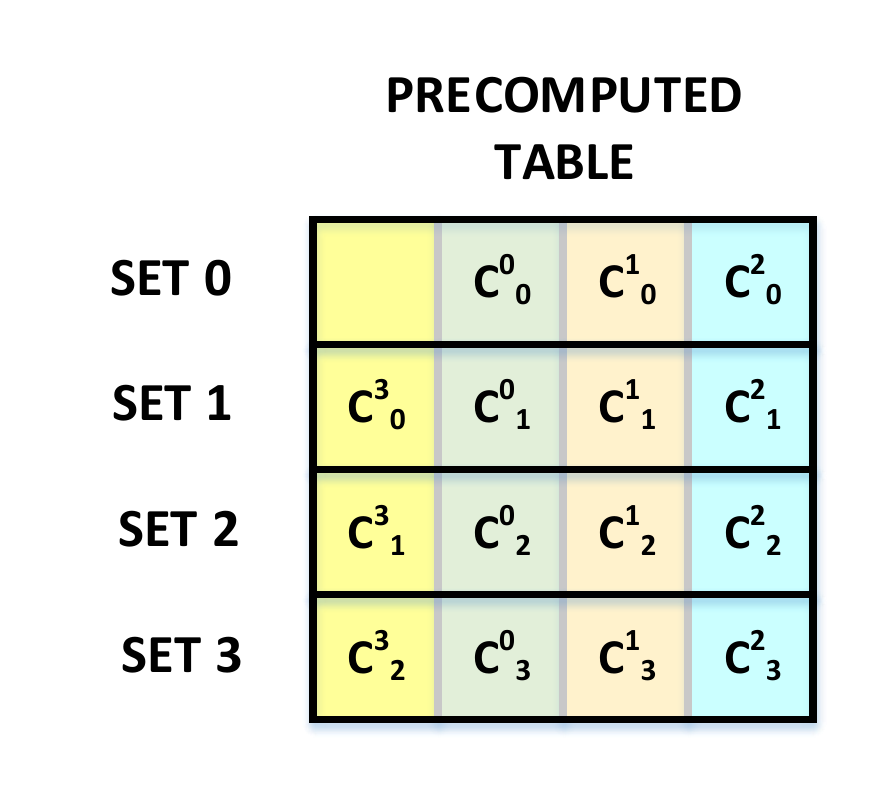}

						\label{fig:non-aligned}}
						
\caption{Cache structure difference for aligned and misaligned precomputed table when using scatter and gather approaches. Observe that when misaligned, an attacker monitoring set 0 can retrieve when $c^3$ is being used.}
\label{fig:al}
\end{figure*}

\subsection{Sliding Window (wolfSSL, mbedTLS, Libgcrypt \& Bouncy Castle)}

WolfSSL and mbedTLS utilize sliding window approaches, which require the window to start with a 1 bit. For these implementations, squaring and multiplications and the pre-computed table were flagged as key dependent. Results for these are shown in Figure~\ref{fig:polandwolf1} and Figure~\ref{fig:polandwolf2}, where instruction leakage is shown in magenta and the instruction + data leakage is shown in blue. We found instruction leakage in wolfSSL due to the usage of the multiplication function, revealing the zeroes between windows. Accesses to the precomputed table also leak information, revealing the window values being used. In the case of mbedTLS we only found table entry accesses to be non-protected. In both cases (easier in the case of wolfSSL), an attacker monitoring the cache would be able to retrieve the key~\cite{inci2016cache,lastlevel}.

\subsection{Fixed Window (OpenSSL and IPP)}\label{sec:fwrsa}
OpenSSL and IPP utilize fixed window approaches to perform modular exponentiations, wich executes a constant number of squares and multiplications for every key value. 
We include the approaches taken by OpenSSL and IPP as examples. The results are presented in Figure~\ref{fig:opandipp1} and ~\ref{fig:opandipp2}. In OpenSSL, we observe that neither the key dependent instructions nor the data leakage is above our noise threshold, i.e., our test categorizes it as non-leaky implementation. One of the reason, among others, is that OpenSSL uses a scatter and gather approach~\cite{He:2007:EGS:1362622.1362684}, i.e., ciphertext powers are stored vertically in the cache and not horizontally. Thus, each cache line contains a little bit of information of each of the ciphertext powers, as seen in Figure~\ref{fig:aligned}. This technique ensures that all the ciphertext powers are loaded whenever one is needed. However, other covert channels might exploit this approach~\cite{q}.

We found no instruction leakage in IPP. In contrast, varying data leakage coming from the pre-computed table was observed for different keys. To better understand this counter-intuitive behavior, we investigated the code. We realized that IPP uses a similar approach to OpenSSL, but the scatter-gather precomputed table is \emph{not} always starting in a cache line boundary, i.e., there is no alignment check when the table was initialized. In consequence, the OS chooses where the table starts with respect to a cache line, i.e., it is not deterministic. Figure~\ref{fig:non-aligned}, in which the table starts at a 1/4 of a cache line, helps to illustrate the problem. An attacker monitoring set 0 obtains cache misses whenever $c^0$, $c^1$ and $c^2$ are utilized, but cache hits whenever $c^3$ is used. Thus, she can detect when window 3 has been utilized by the exponentiation algorithm. If the table is created at the middle of a cache line she can recover when window 2 is being used (since she already knows when 3 is used). Thus, a divide and conquer approach leads to a full key recovery. In contrast, OpenSSL performs an alignment check in the precomputed table. 

\section{Results for ECC}

\begin{figure*}[ht!]
  \centering
	\subfloat[]{
							  \centering
                \includegraphics[width=0.44\linewidth, height = 0.2\textheight, keepaspectratio=true]{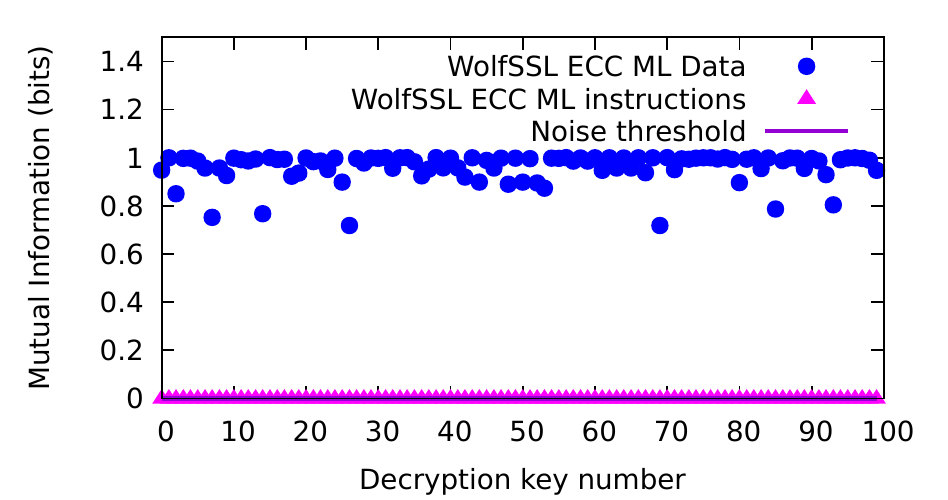}
	
					\label{fig:wecc}}
	\subfloat[]{					
							  \centering
                \includegraphics[width=0.44\linewidth, height = 0.2\textheight, keepaspectratio=true]{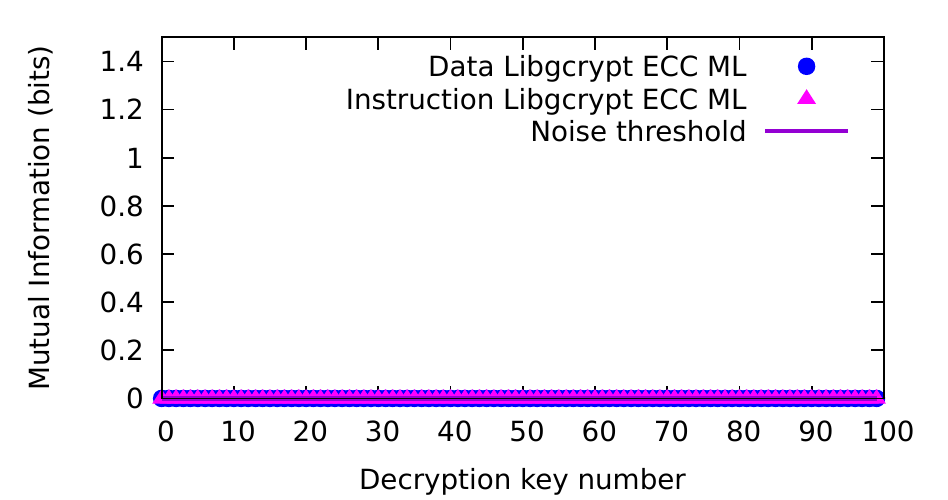}

						\label{fig:libecc}}
						
\caption{Montgomery Ladder ECC leakage for a) wolfSSL and b) Libgcrypt. Magenta indicates instruction leakage while blue indicates instruction+data leakage. Instructions do not give us any information in neither case. Register accesses give us the full key in wolfSSL\vspace{-3ex}}
\label{fig:libandwolf}
\end{figure*}

The following are the four main implementation styles that we encountered for ECC implementation

\subsection{Montgomery Ladder ECC (wolfSSL \& Libgcrypt)}
Only WolfSSL and Libgcrypt implemented this approach. Results are presented in Figure~\ref{fig:libecc}. In the case of Wolfssl, we only found data leakage coming from the key dependent register accesses. Unlike the behavior observed in RSA (in which the cache lines could make the implementation not to leak), the ECC implementation presents leakage for every key. The results for Libgcrypt indicate no presence of leakage because it utilizes a temporary variable on which the computations are performed, and then the contents of it are placed in the corresponding registers in constant execution flow. 

\subsection{Sliding window ECC (wolfSSL)}
\begin{figure}[t!]
  \centering
	\includegraphics[width=0.88\linewidth]{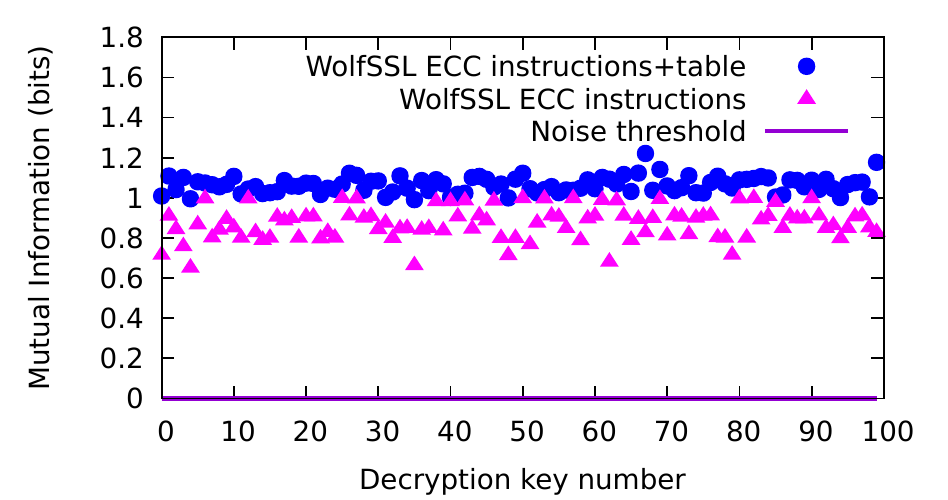}
	\caption{Sliding window ECC leakage in wolfSSL. Magenta indicates instruction leakage, blue indicates instruction+data leakage. Both sources leak information.}
	\label{fig:swolf}
\end{figure}

The default implementation in wolfSSL uses a sliding window approach, very similar to the RSA implementation. The results are shown in Figure~\ref{fig:swolf}, displaying in magenta the instruction leakage and the instruction plus data leakage in blue. We observed that the implementation leaks from both the add instruction, which reveals the zeroes between windows, and from the  accesses to the precomputed table, which reveal the window values. Once again, the ECC implementation of wolfSSL leaks enough information to reveal the secret key.

\subsection{Fixed Window ECC (mbedTLS, Bouncy Castle, NSS)}
\begin{figure*}[t!]
  \centering
	\subfloat[]{
							  \centering
                \includegraphics[width=0.44\linewidth, height = 0.2\textheight, keepaspectratio=true]{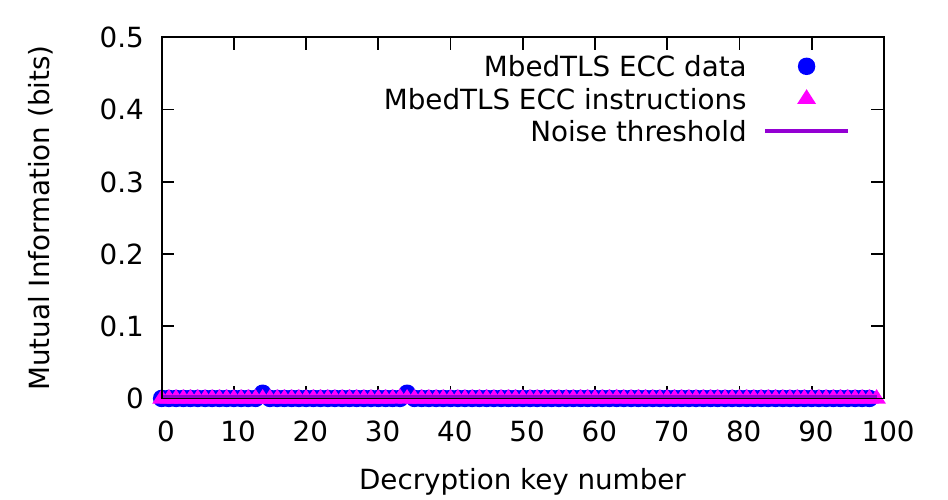}
	
					\label{fig:pecc}}
	\subfloat[]{					
							  \centering
                \includegraphics[width=0.44\linewidth, height = 0.2\textheight, keepaspectratio=true]{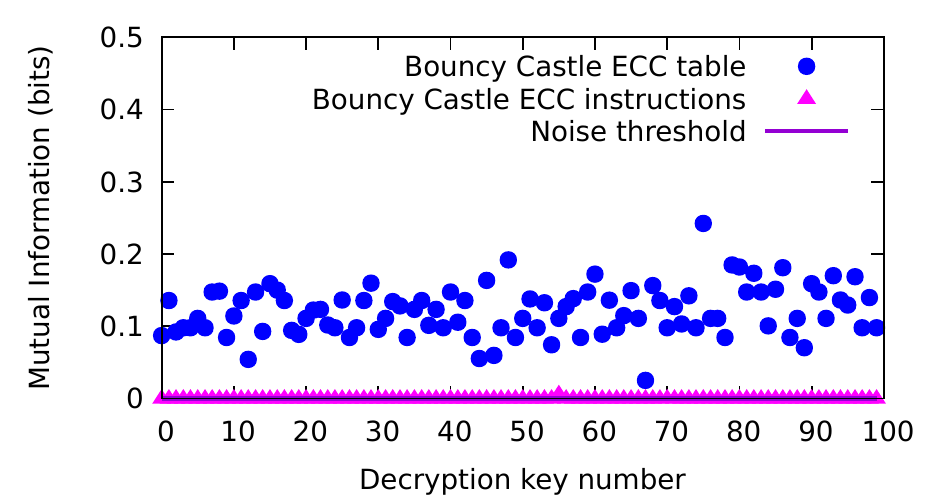}

						\label{fig:becc}}
						
\caption{ECC leakage in mbedTLS and Bouncy Castle. Magenta represents instruction leakage and blue represents instruction+data leakage. MbedTLS does not leak while Bouncy Castle leaks information from the precomputed table}
\label{fig:mbedandbouncy}
\end{figure*}

As with RSA, fixed window implementations process the key in fixed $w$ sized windows. We present two examples, mbedTLS and Bouncy Castle, for which the results are shown in Figures~\ref{fig:pecc} and ~\ref{fig:becc}. Surprisingly, we do not observe any instruction or data leakage in the case of mbedTLS, while we did observe for RSA. One of the differences with respect to the RSA implementation is that mbedTLS accesses all the precomputed table values in a loop before using one, and they choose it in constant execution flow. The fact that their ECC algorithm is well designed but their RSA algorithm is not is surprising. In the case of Bouncy Castle, we do not observe leakage from the instruction side, but we observe leakage in the accesses to the key dependent precomputed table that allows an attacker to recover the key.

\subsection{wNAF ECC (OpenSSL \& LibreSSL)}
OpenSSL and LibreSSL utilize the wNAF scalar multiplication method. To reduce the number of non-zero digits, wNAF works with signed digits, ensuring there are at least $w+1$ zeroes between two non-zero digits. Then the key is processed in a loop, doubling always and performing additions/ subtractions whenever a non-zero digit is found. 
The results for OpenSSL's implementation are shown in Figure~\ref{fig:nafop}, where addition instruction leakage is shown in blue, addition instruction + precomputed table entry leakage is shown in green and the additional sign change instruction leakage effect is shown in magenta. As we can see, the addition instruction already gives us the zero values, the leakage from the precomputed table accesses give us the window value and the sign change function gives us the sign of the window value. Thus, OpenSSL a big door for ECC key recovery attacks.

\section{Leakage summary}\label{sec:sum}

This section summarizes the leakages found in the cryptographic libraries analyzed. The results for all the default (marked in bold) and some non-default implementations
are presented in Table~\ref{tab:1}. The third column indicates whether the implementation leaks the secret or not.  We found that 50\% of the default implementations were susceptible to key recovery attacks (i.e. 12 out of 24).  We believe these numbers are alarming especially given the popularity cache attacks have gained over the last few years. These numbers underline the need for a methodology, like the one proposed in this work, that cryptographic code designers can utilize to detect leakages before the libraries are released to the public.

\section{Comparison with Related Work}

Our work is not the first attempt to detect cache leakage in software binaries. For instance, Zankl et al.~\cite{andi1} use the PIN instrumentation tool to detect instruction leakage in RSA implementations. Demme et al.~\cite{Demme:2012:SVF:2337159.2337172} use simulated cache traces to define a correlation-based side channel vulnerability factor of different hardware characteristics. Almeida et al.~\cite{197207} defined a generic constant time security definition and verifies constant time implementations. Our work differs from them in the following:

\begin{figure}[ht!]
  \centering
	\includegraphics[width=0.88\linewidth]{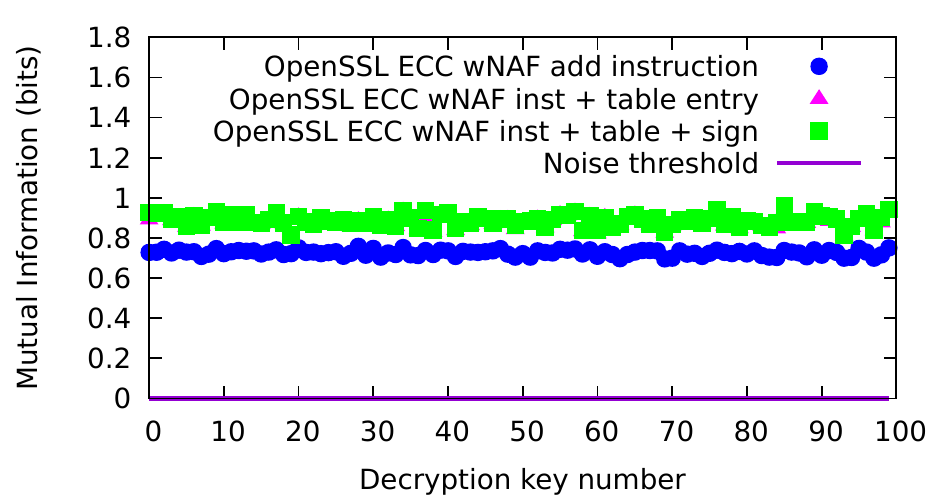}
	\caption{wNAF ECC OpenSSL results. Blue shows addition instruction leakage, magenta adds precomputed table entry leakage and green adds sign change leakage. All possible sources leak information\vspace{-3ex}}
	\label{fig:nafop}
\end{figure}

\begin{packed_itemize}
\item We proposed a sophisticated tool that combines taint analysis with automatic code insertion to track cache leakages of analyzed code while being executed in a realistic environment. We make use of several modern techniques such as the \texttt{flush and reload} technique, taint analysis and mutual information to make the tool efficient and easy to use. 

\begin{table}[ht!]
	\centering
	\caption{Leakage summary for the cryptographic libraries. Default implementations are presented in bold.}
	\label{tab:1}
\resizebox{0.49\textwidth}{!}{\begin{tabular}{|c|c|c|}\hline
	Cryptographic Primitive & Library & Outcome\\
	\hline
	\multirow{9}{*}{AES} & OpenSSL (T-table) &\bf Leaks \\
	& \bf OpenSSL (S-box) & No leak  \\
	& \bf WolfSSL & \bf Leaks \\
	& \bf IPP (v1)\footnotemark[1] & No leak \\
	&  IPP (v2)\footnotemark[1] & No leak \\
	& \bf LibreSSL (S-box) & No leak \\
	& \bf NSS & \bf Leaks \\
	& \bf Libgcrypt & No leak \\
	& \bf Bouncy Castle & \bf Leaks \\
	& \bf MbedTLS & \bf Leaks\\\hline
	 \multirow{9}{*}{RSA} & OpenSSL (Sliding W) & \bf Leaks \\
	& \bf OpenSSL (Fixed W) & No leak \\
	&  WolfSSL (Montgomery L) & \bf Leaks \\
	& \bf WolfSSL (Sliding W) & \bf Leaks \\
	& \bf IPP & \bf Leaks \\
	& \bf LibreSSL  & No leak \\
	& \bf NSS & No leak \\
	& \bf Libgcrypt & No leak \\
	& \bf Bouncy Castle (Sliding W) & \bf Leaks \\
	& \bf MbedTLS (Sliding W) & \bf Leaks \\\hline
	\multirow{8}{*}{ECC} & \bf OpenSSL (WNAF)\footnotemark[2] & \bf Leaks \\
	&  WolfSSL (Montgomery L) & \bf Leaks \\
	& \bf WolfSSL (Sliding W) & \bf Leaks \\
	& \bf IPP\footnotemark[1] & No leak \\
	& \bf LibreSSL  &  \bf Leaks \\
	& \bf NSS & No leak \\
	& \bf Libgcrypt & No leak \\
	& \bf Bouncy Castle (Fixed W) & \bf Leaks \\
	& \bf MbedTLS (Fixed W) & No leak \\\hline
\end{tabular}}
\end{table}
\footnotetext[1]{Due to Intellectual Property conflicts, we are not allowed to disclose the approach taken by the library}
\footnotetext[2]{We verified with OpenSSL that while the 1.0.1 branch leaks, the 1.1.0 branch does not leak}

\item The SVF proposed in~\cite{Demme:2012:SVF:2337159.2337172} uses correlation between an ideal memory-access trace and an cache-observable trace, every time for the entire analyzed code. This means that constant execution flow code will still feature high correlation on leaky architectures, even though the code is not vulnerable. Hence, SVF is a metric for comparing hardware architectures, but cannot be used to find leaking code.

\item Like~\cite{andi1}, our tool looks for dependencies between secret variables and observed leakages to find vulnerabilities. We chose mutual information over correlation, as mutual information is more generic as it does not only look for linear relations and is as easy to implement. Furthermore our work evaluates both instruction and data leakage, while~\cite{andi1} only analyzes instruction leakage, mainly due to the limitation of the proposed PIN based approach to find dynamic memory leakage. 

\item Unlike~\cite{andi1,197207} our methodology takes into account microarchitectural features such as cache line sizes that can affect the code's resistance to cache attacks. This is particularly important to expose leakages caused by OS data cache line missalignments, as those described in Section~\ref{sec:fwrsa} and~\ref{sec:montrsa}, which are not observable without empirical cache traces.

\item Compared to~\cite{andi1,197207}, our methodology detects \emph{exploitable} leakage, as demonstrated by our discovered 12 vulnerabilities that are being fixed. In contrast,~\cite{197207} does not expose new vulnerabilities (it would simply mark all analyzed libraries as non-constant execution flow, i.e., generates lots of false positives) while~\cite{andi1} only exposed one. 
\end{packed_itemize}

\section{Conclusion}
We introduced a proactive tool to detect leakage in security critical code before deployment. The tool uses cache traces for secret dependent memory obtained from execution of the code on the actual microarchitecture and uses the generic Mutual Information Analysis (MIA) as a metric to determine cache leakage. The detection technique can be run across all target platforms without having to redesign it, yet pinpoints parts of code that cause found leakages. 
We used the detection technique to perform the first big scale analysis of popular cryptographic libraries on the three most popular crypto-primitives, namely AES, RSA and ECC. Our results show that about half of the implementations leak information through the LLC. Worse still, a tool like ours can also be used by malicious parties to find exploitable vulnerabilities with ease. This proves the need for cryptographic libraries to resist cache attacks, which can be aided by using the proposed tool in real-life security software development.

\section{Responsible Disclosure}
We disclosed our findings to the cryptographic library designers, resulting thus far in the following CVEs:
\begin{packed_itemize}
\item{\bf IPP:} CVE-2016-XXXX.
\item{\bf WolfSSL:} CVE-2016-XXXX, CVE-2016-XXXX and CVE-2016-XXXX.
\item{\bf BouncyCastle:} CVE-2016-XXXX.
\end{packed_itemize}

The fixes for the remaining issues found in this study will soon be available to the public.

%
%
\section*{Acknowledgments}
This work is supported by the National Science Foundation, under grant No. CNS-1618837.

\bibliographystyle{IEEEtran}
	
\bibliography{mybib}

\end{document}